\documentclass[10pt, sigconf, letterpaper, nonacm]{acmart}
\acmYear{}
\copyrightyear{}
\renewcommand\footnotetextcopyrightpermission[1]{} % removes footnote with conference information in first column
\setcopyright{none}

\def\BibTeX{{\rm B\kern-.05em{\sc i\kern-.025em b}\kern-.08emT\kern-.1667em\lower.7ex\hbox{E}\kern-.125emX}}

\usepackage{subcaption}
\usepackage{float}
\usepackage{multirow}
\usepackage{enumitem}
\usepackage[ruled,noend]{algorithm2e}
\usepackage{url}

\usepackage{breakurl}
\usepackage[normalem]{ulem} 
\setlist[itemize]{leftmargin=*}

\newcommand{\note}[1]{\textcolor{blue}{\textit{#1}}}

\newcommand{\ignoreme}[1]{}

\newcommand{\esther}[1]{\note{Esther: #1}}

\newcommand{\rev}[2]{#2}

\begin{document}

\title[Characterizing Mobile Broadband Access]{A Tale of Three Datasets: Towards Characterizing Mobile Broadband Access in the United States}

\author{Tarun Mangla}
\affiliation{University of Chicago}

\author{Esther Showalter}
\affiliation{
  \institution{University of California, Santa Barbara}
}

\author{Vivek Adarsh}
\affiliation{
  \institution{University of California, Santa Barbara}
}

\author{Kipp Jones}
\affiliation{
  \institution{Skyhook}
}

\author{Morgan Vigil-Hayes}
\affiliation{%
  \institution{Northern Arizona University}
}

\author{Elizabeth Belding}
\affiliation{
  \institution{University of California, Santa Barbara}
}

\author{Ellen Zegura}
\affiliation{%
  \institution{Georgia Institute of Technology}
}

\renewcommand{\shortauthors}{Mangla et al.}

\begin{abstract}
Understanding and improving mobile broadband deployment is critical to bridging the digital divide and targeting future investments. 
Yet accurately mapping mobile coverage is challenging. 
In 2019, the Federal Communications Commission (FCC) released a report on the progress of mobile broadband deployment in the United States~~\cite{FCC2019:Broadband}. This report received a significant amount of criticism with claims that the \rev{LTE} {cellular coverage, mainly available through Long-Term Evolution (LTE)},  was over-reported in some areas, especially those that are rural and/or tribal~\cite{gaoReport}. We evaluate the validity of this criticism using a quantitative analysis of both
the dataset from which the FCC based its report~\cite{fcc477} and a crowdsourced LTE coverage dataset~\cite{skyhook}. \rev{}{Our analysis is focused on the state of New Mexico, a region characterized by diverse mix of demographics-geography and poor broadband access}. We then performed a controlled measurement campaign in northern New Mexico during May 2019. Our findings reveal significant disagreement between the crowdsourced dataset and the FCC dataset regarding the presence of LTE coverage in rural and tribal census blocks, with \rev{disagreement in up to 15\% of these blocks}{the FCC dataset reporting higher coverage than the crowdsourced dataset}. Interestingly, both the FCC and the crowdsourced data report higher coverage compared to our on-the-ground measurements. Based on these findings, we discuss our recommendations for improved LTE coverage measurements, whose importance has only increased in the COVID-19 era of performing work and school from home, \rev{}{especially in rural and tribal areas}.
\end{abstract}

\maketitle
\thispagestyle{empty}
\setcounter{page}{1}
\section{Introduction}
\label{sec:introduction}
Affordable, quality Internet access is critical for full participation in the 21st century economy, education system, and  government~\cite{ Roberts2017:Rural}. % "participation in government access"? Is this wording right? Is 'government access' a term? Shouldn't it just be 'participation in government'? 
Mobile broadband can be achieved through commercial Long-Term Evolution (LTE) cellular networks, which are a proven means of expanding this access~\cite{ITU2017}, but are often concentrated in urban areas and leave economically marginalized and sparsely populated areas underserved~\cite{FCC2019:Broadband}. The U.S. Federal Communication\rev{}{s} Commission (FCC) incentivizes LTE operators serving rural areas~\cite{FCC2017:ConnectAmericaFund, prieger2017} and maintains transparency by releasing maps from each operator showing geographic areas of coverage~\cite{fccLTEData}. Recently third parties have challenged the veracity of these maps, claiming these maps over-represent true coverage, and
thus may discourage much-needed investments.

Most of these claims, however, are either focused on limited areas where a few dedicated researchers can collect controlled coverage measurements (e.g., through wardriving), or are mainly qualitative in nature~\cite{ms2019:ruralbroadband,challengesRWA, RWA2018tmob}. \emph{As dependence on mobile broadband connectivity increases, especially in the face of the COVID-19 pandemic, mechanisms that quantitatively validate FCC coverage datasets at scale are becoming acutely necessary to evaluate and direct resources in Internet access deployment efforts~\cite{Pew2019:Mobile, lutu2020characterization}.} This is an issue of technology 
and technology policy, with equity and fairness implications for society.

An increasingly widespread approach to measure coverage at scale is through crowdsourcing wherein users of the LTE network contribute to coverage measurements. The FCC has recently advocated for the use of crowdsourcing to validate coverage data reported by operators~\cite{fccRecommendations}. In this context, we take a data-driven, empirical approach in this work, comparing coverage from a representative crowdsourced dataset with the FCC data. More specifically, our analysis is guided by the following questions: (i) How consistent are existing LTE coverage datasets, ii)  where and how do their coverage estimations differ, and what trends are present?

We specifically consider a crowdsourced coverage estimate from Skyhook, a commercial location service provider \rev{}{that uses a variety of positioning tools to offer precise geolocation}. 
We select Skyhook because it crowdsources \rev{}{cellular coverage} measurements from end-user applications that subscribe to its location services. Such \textit{incidental} crowdsourcing can potentially provide richer coverage data compared to a \textit{voluntary} form of crowdsourcing where a user has to explicitly commit to contributing coverage data. We examine this by comparing the Skyhook measurements with those of OpenCellID, an open but voluntary crowdsourced dataset~\cite{opencellid}. As will be shown in Section~\ref{sec:comparison_crowdsourcing}, we find that the density of the crowdsourced datasets varies significantly by the methodology of data collection, especially in rural areas. In the regions we studied, incidental crowdsourcing (Skyhook) gathered up to $11.1$x more cell IDs than voluntary crowdsourcing (OpenCellID). 

Using Skyhook as an extensive crowdsourced dataset, we quantify how widely and where the crowdsourced coverage data differs from the FCC data. We specifically focus on the state of New Mexico\footnote{Our methodology is not specific to New Mexico and can be easily extended to other regions in the U.S.}, selected for its mix of demographics, diverse geographic landscape, and our partnership with community stakeholders within the state. We compare coverage at the level of census blocks\footnote{\rev{}{We use the FCC methodology wherein a census block is considered covered if the centroid is covered~\cite{centroidMethodology}}} which are further grouped into urban, rural, and tribal\footnote{Tribal areas have consistently experienced the lowest broadband coverage rates in the United States for the past decade~\cite{FCC2019:Broadband}} categories. We find that the FCC and Skyhook LTE datasets have a disagreement as great as $15\%$ in rural census blocks with the data from FCC claiming higher coverage than Skyhook. A major concern in interpreting this comparison is accounting for coverage disagreement as a result of lack of data points in the crowdsourced dataset. To confirm the availability of users to provide data points, we check for the presence of alternate cellular technologies (e.g., 2G or 3G) within these census blocks and observe a significant number (up to $9\%$ in tribal rural areas) where such alternates are present, providing evidence that users do visit those blocks but cannot access LTE. These results, \rev{}{similar to a recent study on fixed broadband~\cite{major2020no}}, suggest a need for incorporating mechanisms to validate the operator-submitted data into the FCC's LTE access measurement methodology, especially in  rural and tribal areas. 

Finally, we compare both FCC and Skyhook coverage maps to our own controlled coverage measurements collected from a northern section of New Mexico. Interestingly, we find that both FCC and Skyhook datasets report higher coverage relative to our controlled measurements with the former showing a higher degree (by up to 26.7\%) of over-reporting than the latter. 
Understanding the causes of these inconsistencies is important for effectively using crowdsourced data to measure LTE coverage, especially as crowdsourcing is increasingly viewed as preferable
to provider reports. We conclude with recommendations for improving LTE coverage
measurements, whose importance has only increased in the COVID-19 era of performing work 
and school from home. 
\begin{table}[t!]
    \centering
    \small
    \begin{tabular}{|l|c|c|c|}
        \hline
        \textbf{Data Set} & \textbf{Points of} & \textbf{Format}  & \textbf{Methodology} \\
                          & \textbf{Collection}           &                  &                      \\
        \hline
        \hline
         FCC        & Polygon        & Shapefile   & Operator-reported  \\ 
                    & overlay        &             & with Form 477   \\\hline
         Skyhook    & Cell signal    & CSV         & Incidental  \\     
                    & point          &             & crowdsourcing   \\\hline
         Author Controlled    & Cell signal    & CSV        & Wardriving \\
         Measurements & point         &            & \\
         \hline
    \end{tabular}
    \caption{Summary of coverage data sets.}
    \label{tab:data_overview}
    \vspace{-2.5em}
\end{table}

\section{Background and Datasets} 
\label{sec:dataset}

\rev{In this section we describe the  LTE coverage datasets compared in this analysis.}{In this section, we first provide an overview of the LTE network architecture. This is followed by a description of the LTE coverage datasets compared in our analysis.} These datasets are  summarized in Table~\ref{tab:data_overview}. We also note the limitations associated with each data collection methodology.

\begin{figure}[t]
\centering
\begin{minipage}[t]{0.23\textwidth}
\centering
	\includegraphics[width=\textwidth, keepaspectratio]{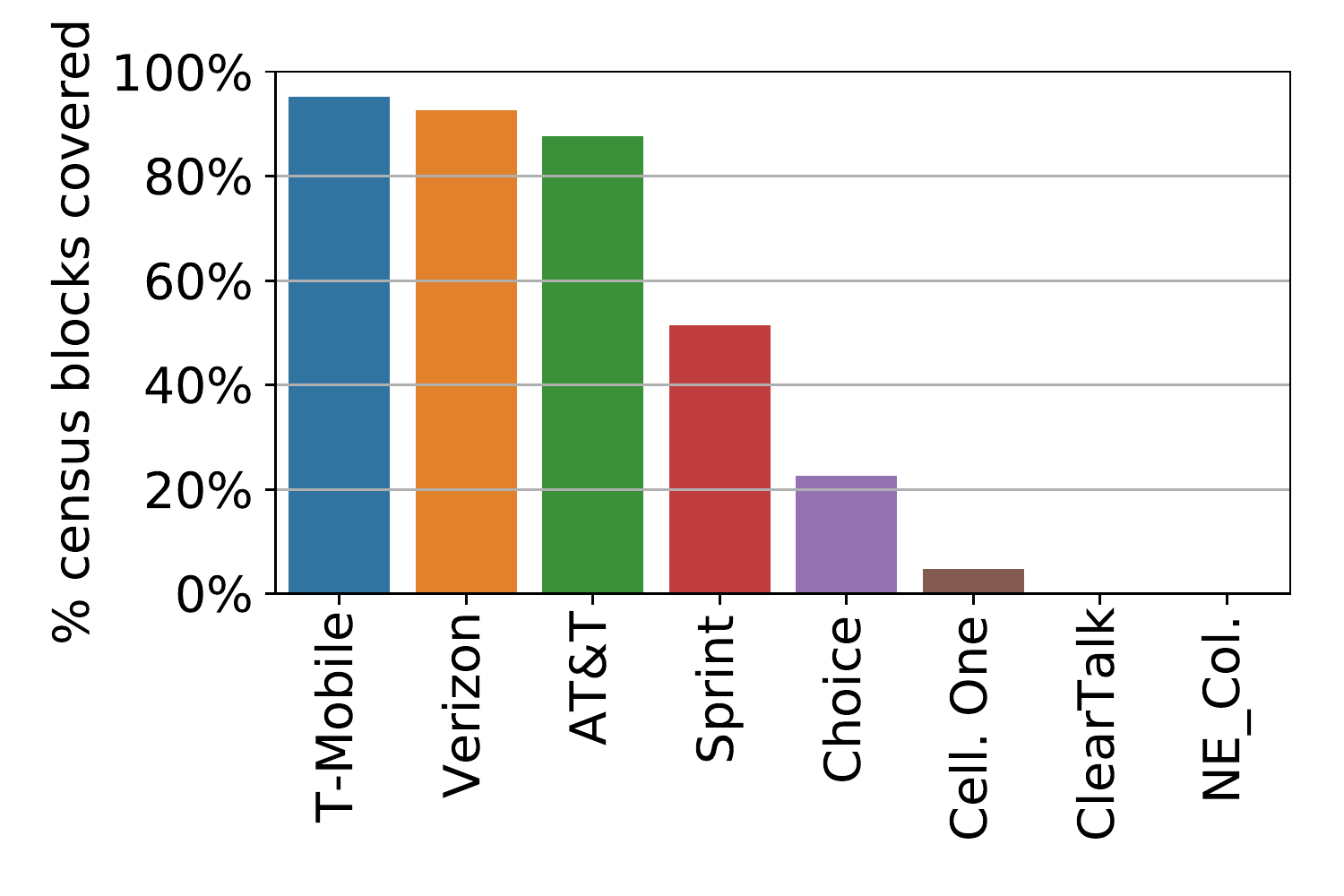}
	\caption{\rev{}{LTE operators by census block coverage based on FCC data.}}
	\label{fig:fcc_coverage}
\end{minipage} \hfill
\begin{minipage}[t]{.235\textwidth}
  \centering
	\includegraphics[width=\textwidth,keepaspectratio]{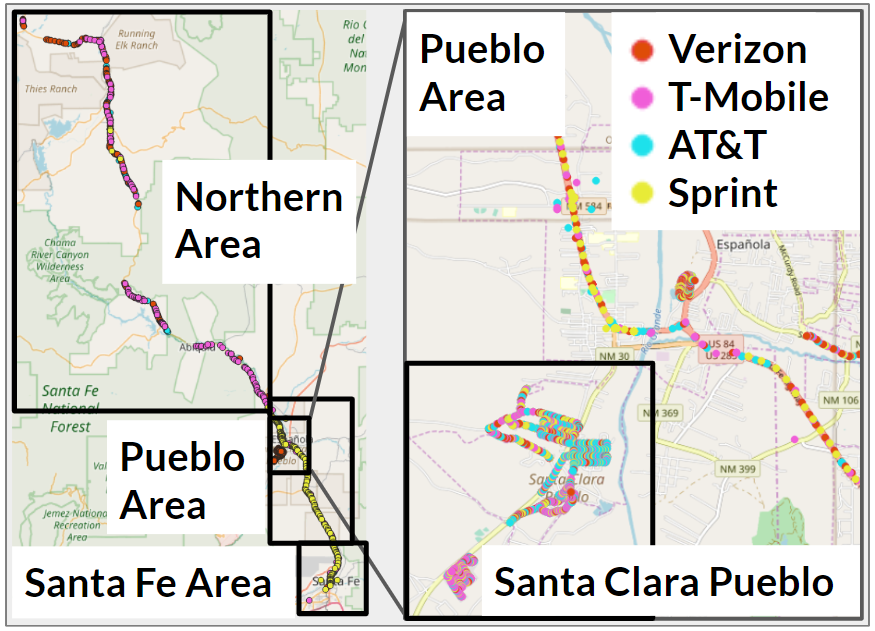}
	\caption{Map of author wardriving areas in New Mexico.}
	\label{fig: ue_drive}
\end{minipage}
\end{figure}

\subsection{LTE Network Architecture}
\rev{}{
Internet access in an LTE network is available through base stations (known as eNodeBs) operated by the network provider. User equipment (UE), such as smartphones, tablets, or LTE modems,  connects to the eNodeB over the radio link. The eNodeB is connected to a centralized cellular core known as the Evolved Packet Core (EPC). This connection is  typically through a wired link forming a middle-mile connection. The EPC consists of several network elements including a Packet Data Network Gateway (PGW), which is the connecting node between an end-user device  and the public Internet. Thus, LTE broadband access depends on multiple factors including radio coverage, middle-mile capacity, and interconnection links with other networks (e.g., transit providers, content providers) in the public Internet. However, the focus of this article is on understanding the last-mile LTE connectivity characterized by the radio coverage of the eNodeB.}

\rev{}{An eNodeB controls a single cell site and consists of several radio transceivers or cells mounted on a raised structure such as a mast or a tower. The radio cells use directional antennas, where each antenna provides coverage in a smaller geographical area using one frequency band. The radio cells can be identified through a globally unique number called cell identifier (or cell ID), which is also visible to an end-user device in range of the cell. The cell ID enables aggregation of connectivity and signal strength information from multiple UEs connected to the same cell, which can then be used to estimate the geolocation of a cell along with its coverage (see Section~\ref{sec:skyhook_data}).}

\subsection{FCC Dataset}
\label{sec: fcc_data}

The FCC LTE broadband dataset consists of coverage maps in shapefile format that depict geospatial LTE network deployment for each cellular operators in the U.S. The FCC compiles this dataset semi-annually from operators through Form~477. Every operator that owns cellular network facilities must participate in this data collection. The operators submit shapefiles containing detailed network information in the form of geo-polygons along with the frequency band used in the polygon and the minimum advertised upload and download speeds. The methodology used for obtaining these polygons is proprietary to each operator. Ultimately, the FCC publishes only a coverage map that represents coverage as a binary indicator: in any location, cellular service is either available though an operator, or it is not. 

\rev{We use the latest binary coverage shapefiles, available on the FCC's website, from December 31, 2018~\cite{fcc477}}{We use the binary coverage shapefiles, available on the FCC's website, from June 2019\footnote{At the time of this analysis, data from December 2019 was also available on the FCC website. However, we use data from June 2019 as the other two datasets in our analysis are collected around this period.}}. Figure~\ref{fig:fcc_coverage} shows the eight LTE network operators present in the state of New Mexico (NM) and the percentage of total census blocks in NM covered by each operator. Note that \rev{}{we use one of the FCC methodologies to report mobile broadband access, wherein a census block is considered covered} if the centroid of the census block is covered~\cite{centroidMethodology}. In this paper, we limit our analysis to the top four cellular operators due to their significantly greater prevalence in NM; these operators are also the top four cellular operators in the United States more broadly.

\noindent\emph{Limitations:} These coverage maps are generated using \rev{}{predictive} models that are proprietary to the operator~\cite{gaoReport} and not generally reproducible. Furthermore, the publicly available dataset consists of binary coverage and \rev{the FCC does not provide performance-related data.}{lacks any performance-related data.\footnote{The FCC has only recently (beginning December 2019) started providing speed data along with coverage information.}}

\subsection{Skyhook Dataset} \label{sec:skyhook_data}
Skyhook is a location service provider that uses a variety of positioning tools, including a database of cell locations, to offer precise geolocation to subscribed applications~\cite{skyhook}. Through apps that subscribe to Skyhook's location services, user devices report back network information, which is gathered into anonymous logs and used to further improve the localization service. Through a data access agreement we are able to view the \textit{cell location database} consisting of a list of unique cell IDs along with the cell technology (e.g., 3G vs LTE), estimated location, and the estimated coverage. The database was originally constructed through extensive wardriving but is now managed and updated using measurements gathered by devices using the Skyhook API for localization. The device measurements with the same cell ID are combined to estimate the cell location and coverage in the following manner: 

\noindent\textbf{Cell location estimation}: A grid-based methodology similar to that proposed by Nurmi et al.~\cite{nurmi2010grid} is used to estimate the cell tower location.  % \tarun{check if there is any patent about the Skyhook's cell positioning technology that we can cite}. 
Specifically, Skyhook divides the geographic area into $7$ m squares and groups measurements in the same square to obtain a central measure of the square's signal strength. This is done to reduce the bias due to large numbers of measurements coming from the same area (e.g. a popular gathering place). A weighted average of the signal strength is then used to estimate the cell location.  

\noindent\textbf{Estimation of cell coverage radius}: Skyhook also provides an estimate of the cell's coverage radius using a proprietary method based on the path-loss gradient~\cite{tse2005fundamentals}. \rev{}{The path-loss gradient 
approximates how the wireless signal attenuates as a function of the distance from the transmitter (radio cell in this case).} The value of the path-loss gradient depends on several factors such as environment (foliage, buildings), geographic topography, and cell signal frequency. Skyhook estimates the path-loss gradient using field observations of cell signal strength readings along with their distributed geographic locations. Ideally, the \rev{path loss}{signal attenuation} varies based on the direction and the distance from the cell. However, to reduce the complexity of coverage estimation, Skyhook's cell coverage estimation heuristic calculates only one path-loss gradient for a single cell. The path-loss gradient is then used in a set of parameterized equations to estimate the cell coverage radius. The parameters in these equations have been determined with careful research and testing over more than 10 years.

The cell location database is updated regularly with recalculation of cell location and cell coverage radius using the new device measurements that have been collected since the last update. For our analysis, we use the cell location database last updated on \rev{April 30}{June 10}, 2019.  

\noindent\emph{Limitations:} Since database entries are crowdsourced when the device passes within range of a cell, this dataset is more comprehensive in population centers and highways where people more often occupy. If there are too few measurements overall, or if measurements are primarily sourced from the same grid section, then the cell location estimate can be inaccurate.

\subsection{Targeted Measurement Campaign}
\label{subsec:active_measurements}
To complement these datasets, we performed a targeted measurement campaign collecting coverage information through 120 miles of Rio Arriba county in New Mexico over a period of five days beginning May 28, 2019. Figure~\ref{fig: ue_drive} shows the locations of ground measurements and the four descriptive area labels we use for this analysis. The North area measurements were taken on highways passing primarily through national forest. The Pueblo area measurements were taken from highways within tribal jurisdiction boundaries. In Santa Clara Pueblo, tribal leadership permitted us to collect additional measurements in residential zones. Finally, the Santa Fe area consists of highway measurements between the pueblos and downtown Santa Fe.
While limited in scale, these active measurements provide an important comparison point for coverage and user experience. As described in Section~\ref{sec:introduction}, we selected these areas of New Mexico for their mix of tribal and non-tribal demographics; tribal lands tend to have the highest coverage over-statements and the most limited cellular availability within the United States~\cite{FCC2019:Broadband}.

Our measurements consist of {\em service state} and signal strength readings recorded on four Motorola G7 Power (XT1955-5) phones running Android Pie (9.0.0). {\em Service State} is a discrete variable indicating whether the phone is connected to a cell.  Measurements were collected using the \emph{Network Monitor} application~\cite{networkmonitor}. An external GlobalSat BU-353-S4 GPS connected to an Ubuntu Lenovo ThinkPad laptop gathered geolocation tags that were matched to network measurements by timestamp. Each phone was outfitted with a SIM card from one of the four top cellular operators in the area:  Verizon, T-Mobile, AT\&T, and Sprint. The phones recorded service state and signal strength every 10 seconds while we drove at highway speeds (between $40$ and $65$ miles per hour) in most places and less than $10$ miles per hour in residential areas (Santa Clara Pueblo). 

\noindent\emph{Limitations:} 
Our wardriving campaign was intensive in terms of human effort, economic cost, and time, making it difficult to scale. The dataset does not capture any temporal variations in coverage as the measurements were collected over a short span of time. It is possible that driving
speed or device configuration affects the measurements, e.g., indicating no coverage when a stationary measurement might have detected coverage~\cite{fida2018impact}. We have no evidence that this occurred, but it might warrant some additional investigation. 

\section{Analysis}
\label{sec:analysis}
In this section, we first evaluate of Skyhook as a representative crowdsourced dataset by comparing it with a popular \textit{voluntary} crowdsourced data from OpenCellID~\cite{opencellid}. This is followed by comparison of coverage across the FCC, Skyhook, and our wardriving measurement data. Our comparison is guided by the following questions:  (i) what is the degree of coverage agreement across the datasets, ii) where and how do their coverage estimations differ?

\begin{table*}[t!]
%\vspace{-3mm} 
  %  \centering
  \small
\begin{tabular}{|c|c|c|c|c|c|c|c|c|c|}
    \hline
   \multirow{2}{*}{\begin{tabular}[c]{@{}c@{}} \textbf{County} \\  \textbf{classification} \end{tabular}} & 
   \textbf{Region } 
   & \textbf{County }
   &  \multirow{2}{*}{\begin{tabular}[c]{@{}c@{}} \textbf{Population} \\  \textbf{density (per sq. mile)} \end{tabular}} & \multicolumn{2}{c|}{\textbf{Skyhook}} & \multicolumn{2}{c|}{\textbf{OpenCellID}} &  \multirow{2}{*}{\begin{tabular}[c]{@{}c@{}} \textbf{Common}\\  \textbf{CIDs} \end{tabular}} \\
   & & \textbf{Name}& & \textbf{CIDs (\#)}  & \textbf{\% Overlap} & \textbf{CIDs (\#)}  & \textbf{\% Overlap} & \textbf{CIDs} \\
    \hline\hline
                  & Western & Los Angeles, CA  & 2,490.3    &  133,484  &  28\%  &  39,875  &  92\%  &  36,816  \\ 
    Large Metro   & Central & Denver, CO       & 4,683.0    &  11,061   &  24\%  &  3,136   &  86\%  &  2,689  \\ 
                  & Eastern & Fulton, GA       & 1,994.0    &  27,809   &  22\%  &  7,225   &  86\%  &  6,194  \\ \hline
                  & Western & Imperial, CA     & 43.5       &  1,818    &  17\%  &  336     &  93\%  &    311  \\  
    Small Metro   & Central & Do{\~n}a Ana, NM & 57.1       &  1,870    &  32\%  &  663     &  89\%  &    592  \\ 
                  & Eastern & Bibb, GA         & 613.0      &  1,953    &  21\%  &  464     &  89\%  &    413  \\  \hline
                  & Western & Tehama, CA       & 21.7       &  733      &  17\%  &  158     &  80\%  &    126  \\  
    Micropolitan  & Central & Rio Arriba, NM   & 6.7        &  333      &  8\%   &  30      &  87\%  &     26  \\   
                  & Eastern & Pierce, GA       & 61.3       &  164      &  9\%   &  21      &  67\%  &     14  \\    \hline
\end{tabular}
  \caption{Characteristics and cell ID (CID) counts in selected counties.}
\label{tab: county_stats}
%\vspace{-3.0mm}
\end{table*}

\subsection{Comparison of Crowdsourced Datasets}
\label{sec:comparison_crowdsourcing}

We compare the Skyhook dataset with a publicly available crowdsourced dataset -- OpenCellID. Unwired Lab's OpenCellID\footnote{OpenCellID Project is licensed under a Creative Commons Attribution-ShareAlike 4.0 International License. %/esther{website asks to mention this when citing}
} project provides a publicly available dataset of cell IDs along with their estimated location. The dataset is derived from crowdsourced UE signal strength measurements similar to Skyhook. However, the UE measurements in this case come from users voluntarily installing the OpenCellID application on their smartphone~\cite{opencellid} and manually choosing what data to upload. We differentiate this \textit{voluntary} crowdsourcing method of data collection from Skyhook's \textit{incidental} crowdsourcing method, where users of the Skyhook API contribute to the data by default. We specifically compare the number of unique LTE cells and the recentness of the measurements in both datasets. We consider each of these factors to contribute to the overall density of the dataset. 

\begin{figure}[t]
\centering
    \includegraphics[width=0.25\textwidth,keepaspectratio]{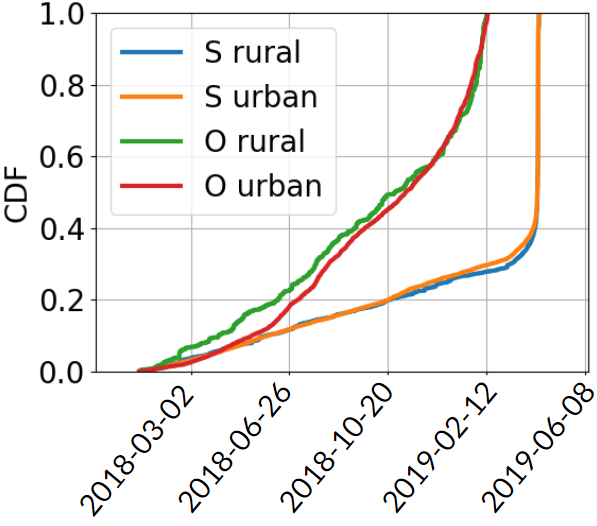}
	\caption{CDF of cell updates in Skyhook (S) and OpenCellID (O).}
	\label{fig: s_o_times}
	\vspace{-2em}
\end{figure}

\noindent\textbf{Methodology}: While our coverage comparison will be focused on New Mexico, we analyze our selected crowdsourced data more broadly by considering these datasets within a set of counties of differing population densities across the United States. The counties are selected from three  areas of the United States:  Western (California), Central (New Mexico and Colorado), and Eastern (Georgia). Within each region, we consider three different kinds of counties as defined by the National Center for Health Statistics' 2013 Urban-Rural Classification Guide \cite{nchs2013}. These are: (i) \emph{large metropolitan} (large), which contain a population of at least one million and a principle city; (ii) \emph{small metropolitan} (small), which contain a population of less than 250,000; and (iii) \emph{micropolitan} (micro), which must have at least one urban cluster of at least 10,000, but a total population of less than 50,000. This enables us to study differences based on population density and geographic region for the crowdsourced datasets. We select three counties of each population category, for a total of nine counties, to compare these two datasets. We describe these counties in Table~\ref{tab: county_stats}. For each county, we show the 2018 population density estimated from the U.S. Census Bureau's 2010 census records~\cite{pop2010}. We first count the number of unique cell IDs that appear in both datasets for each county, as shown in Table~\ref{tab: county_stats}. The ``\% Overlap" column in Table~\ref{tab: county_stats} shows the percentage of each dataset's cell IDs that also appear in the other dataset, and the ``Common CIDs" column shows the exact number of common cell IDs.

\noindent\textbf{Results}:  Overall, Skyhook reports a greater number of cells (2.8x - 11.1x) for all counties.  The difference is particularly pronounced in micro counties. This suggests that relying on volunteers to download an application and offer network measurements may not be the most accurate method for assessing LTE coverage in rural areas. Furthermore, Skyhook includes a majority of the cells that appear in OpenCellID.

We next consider how recently each cell ID record was updated with a new measurement. Figure~\ref{fig: s_o_times} shows the CDF of the latest measurement date for cells in both the datasets, where cells are split into those located in urban and rural census blocks. Almost 60\% of the cells in Skyhook were last updated in the month of June 2019, but the most recent update in OpenCellID was in February 2019. Furthermore, cells in rural census blocks were updated less recently than urban census blocks in OpenCellID, while the difference is negligible in the Skyhook dataset. This suggests that the Skyhook dataset is updated more regularly than OpenCellID, thus making it more likely to represent any changes in the network infrastructure.

\subsection{Comparison of Coverage}
\label{sec:analysis_coverage_comparison}

\begin{table*}[t]
\small
\begin{tabular}{|c|c|c|c|c|c|c|c|c|c|}
\hline
\multirow{2}{*}{\begin{tabular}[c]{@{}c@{}}\textbf{Census}\\   \textbf{block type}\end{tabular}} & \multirow{2}{*}{\begin{tabular}[c]{@{}c@{}}\textbf{Total census} \\ \textbf{blocks}\end{tabular}} & \multicolumn{2}{c|}{\textbf{Verizon}} & \multicolumn{2}{c|}{\textbf{T-Mobile}} & \multicolumn{2}{c|}{\textbf{AT\&T}} & \multicolumn{2}{c|}{\textbf{Sprint}} \\ \cline{3-10} 
& & \textbf{FCC} & \textbf{Skyhook} & \textbf{FCC} & \textbf{Skyhook}  & \textbf{FCC}    & \textbf{Skyhook}     & \textbf{FCC}     & \textbf{Skyhook} \\ \hline \hline
\tiny{Non-Tribal Rural}& 93,680     & 89\%  & 77\% & 94\%  & 86\%  & 85\% & 79\%     & 39\% & 49\% \\ \hline
\tiny{Non-Tribal Urban}& 41,872     & 100\%  & 100\%     & 100\%  & 100\% & 99\% & 99\%     & 96\% & 99\% \\ \hline
\tiny{Tribal Rural}     & 30,588     & 93\%  & 80\% & 92\%  & 63\%  & 78\% & 73\%     & 27\% & 41\% \\ \hline
\tiny{Tribal Urban}     & 2,469 & 100\% & 99\% & 95\%  & 94\%  & 93\% & 94\%     & 75\% & 88\% \\ \hline
All & 168,609    & 93\%  & 84\% & 95\%  & 85\%  & 88\% & 83\%     & 52\% & 61\% \\ \hline
\end{tabular}
\caption{\rev{}{Percentage of total census blocks covered according to FCC and Skyhook.}}
\label{tab: coverage_btype}
\vspace{-3mm}
\end{table*}

\ignoreme{
We  compare LTE coverage across three datasets: FCC, Skyhook, and active measurements. While our focus is on the state of New Mexico % as our targeted measurement area 
due to its diverse geographic landscapes and political jurisdictions~\cite{NMEDD2019:Climate}, nothing in our methodology is particular to this state, except
our ability to conduct active measurements there. 
}

\subsubsection{Coverage comparison between the FCC and Skyhook}
\label{subsubsec:coverage_s_f}
We first compare a coverage shapefile generated from Skyhook cell locations and estimated coverage ranges with the FCC map for each operator.

 \begin{table}[]
\small
\begin{tabular}{|c|c|c|c|c|c|}
\hline
\begin{tabular}[c]{@{}c@{}}\textbf{Block}\\  \textbf{type}\end{tabular} & \begin{tabular}[c]{@{}c@{}}\textbf{Total} \\\textbf{ blocks}\end{tabular} & \textbf{Verizon} & \textbf{T-Mobile} & \textbf{AT\&T} & \textbf{Sprint} \\ \hline \hline
\tiny{Non-Tribal Rural} & 93,680 & 14,013 & 9,025 & 8,705 & 1,355 \\ \hline
\tiny{Non-Tribal Urban} & 41,872 & 0     & 0    & 213  & 25   \\ \hline
\tiny{Tribal Rural}  & 30,588 & 5,109  & 9,150 & 3,004 & 230  \\ \hline
\tiny{Tribal Urban}  & 2,469  & 4     & 14    & 4    & 0    \\ \hline
\end{tabular}
\caption{\rev{}{Number of census blocks where there is coverage according to FCC but no coverage according to Skyhook.}}
\label{tab: NCSH_CFCC}
\vspace{-3mm}
\end{table}

\noindent\textbf{Methodology}: 
We consider coverage at the census block level for this comparison. In addition to reporting coverage shapefiles, the FCC reports coverage at a census block level and considers a census block as covered if the centroid of the census block falls within a covered region~\cite{centroidMethodology}. We generate a similar census block level coverage map per-operator using Skyhook's estimated coverage. To do so, we first obtain the coverage shapefile for each operator using a cell's estimated location and coverage radius. Then we use the FCC centroid methodology to generate the Skyhook LTE coverage map at the census block level. We use the Python GeoPandas 0.8.2 library for the associated spatial operations~\cite{geopandas}.
We group census blocks into four categories: Non-Tribal Urban, Non-Tribal Rural, Tribal Urban, and Tribal Rural. This is done to explore whether the degree of agreement of the two datasets varies across these dimensions. 
We use the U.S. Census Bureau's classification of urban and rural blocks and its boundary definitions of tribal jurisdiction for this categorization~\cite{urbanRural}. 
In this analysis we consider census blocks as tribal if they overlap with any tribal boundaries. \rev{}{We varied the tribal labeling schemes such as classifying a census block tribal if the centroid of the block is within a tribal boundary. However, the results remain qualitatively similar and do not impact the findings presented here.}

\noindent\textbf{Results}: 
Table~\ref{tab: coverage_btype} shows the percentage of total census blocks covered by each cellular operator, according to the FCC and Skyhook data, broken down by census block type. Among the four operators, T-Mobile covers the greatest number  of census blocks based on both FCC and Skyhook data, while Sprint covers the fewest. All  four cellular operators have relatively higher coverage for both tribal and non-tribal urban census blocks. However, all operators except Verizon offer their lowest coverage in tribal rural areas. For some operators, the differences between non-tribal rural and tribal rural are as great as \rev{$23\%$}{$23\%$ (based on Skyhook data) and $11\%$ (based on FCC data)}.

\begin{figure}[t]
%	\vspace{-2mm}
    \begin{subfigure}{0.22\textwidth}      
    		\centering
        \includegraphics[width=\textwidth,keepaspectratio]{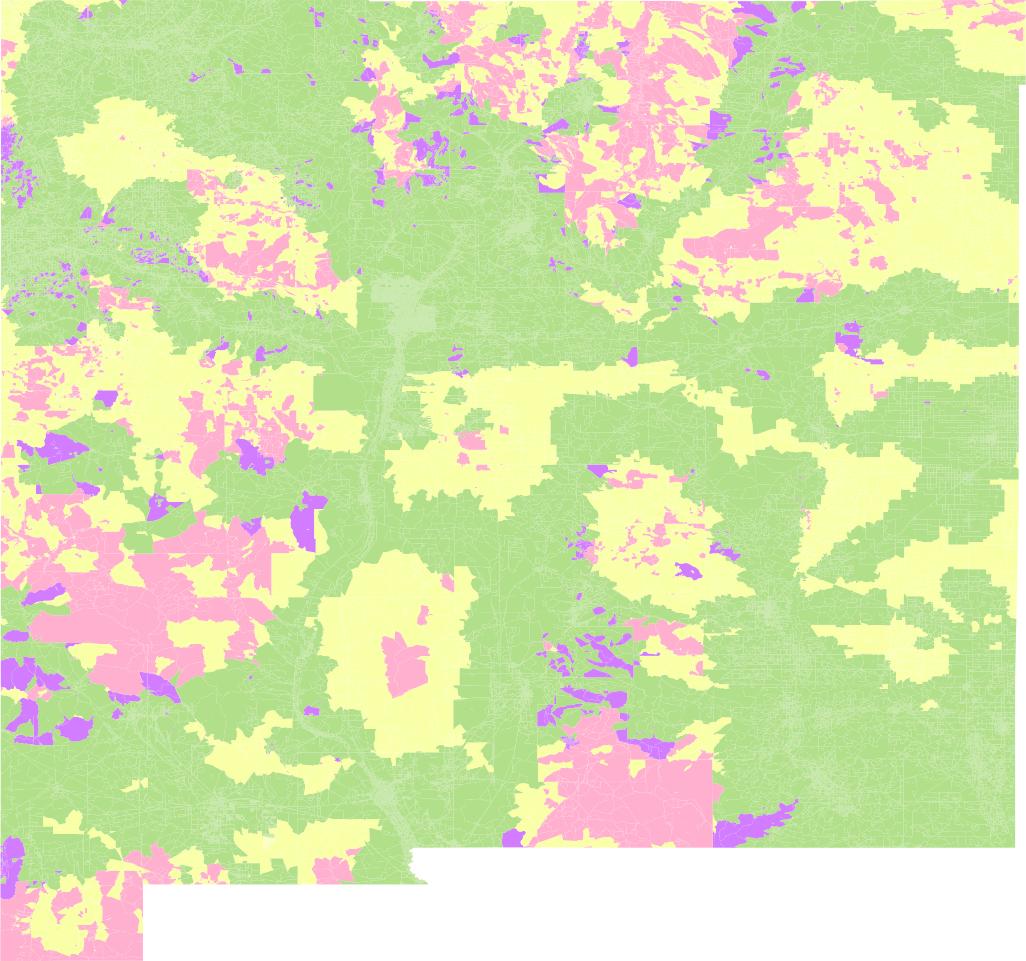}
        \caption{Verizon}
 		\label{subfig:verizon_combined}
    \end{subfigure}%
    \hfill
	\begin{subfigure}{0.22\textwidth}   
        \centering
        \includegraphics[width=\textwidth]{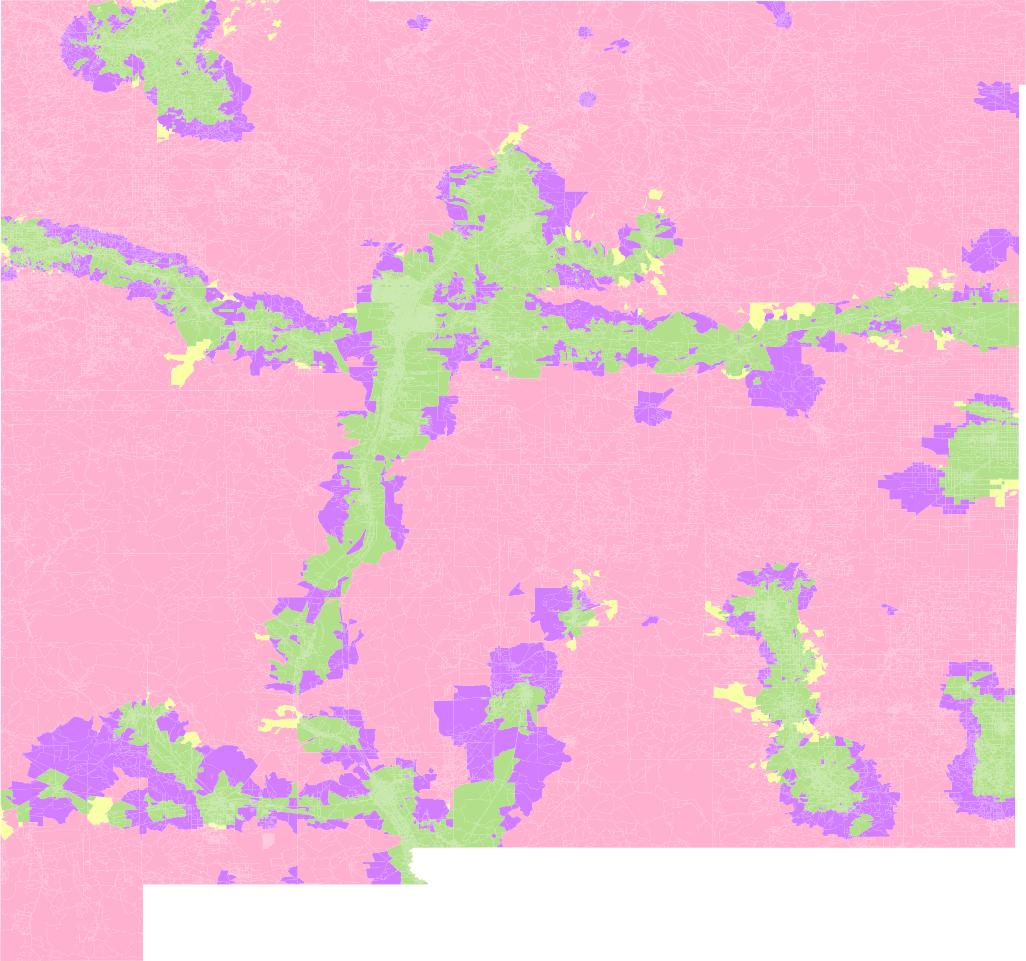}
    \caption{Sprint}
    \label{subfig:sprint_combined}
    \end{subfigure}% 
     \vspace{-3mm}
	\caption{\rev{}{Comparison of LTE coverage maps of New Mexico. Yellow blocks are covered in the FCC map but not in Skyhook; purple blocks are covered in the Skyhook map but not the FCC. Green blocks are covered in both, and pink blocks are covered in neither.}}
	\label{fig:VScoverage_map}
	\vspace{-1em}
\end{figure}

The extent of LTE coverage differs between the two datasets. For three out of four providers, Skyhook shows lower coverage than the FCC, particularly in the rural census blocks. For instance, the FCC T-Mobile data shows coverage in \rev{$90\%$}{$92\%$} of tribal rural blocks, whereas Skyhook shows coverage in only $63\%$ of such blocks. On the other hand, Skyhook shows a higher number of census blocks covered than the FCC for Sprint. \rev{We believe this could be because the Skyhook coverage map data was updated in June 2019, whereas the FCC coverage maps correspond to December 2018. Network operators could have expanded service in that time period.}{The higher coverage in the case of Sprint could have been due to multiple reasons, including: (i) there are differences in the propagation models used by Skyhook and Sprint to estimate coverage with the former's models being more generous than the latter's, and (ii) the Skyhook data is collected across time and Sprint may have discontinued or temporarily disabled some of the cells, which is challenging to detect from the crowdsourced data.}

Figure~\ref{fig:VScoverage_map} visually compares the LTE coverage maps from the FCC and the Skyhook datasets for Verizon and Sprint. We more deeply examine the  discrepancy mapped in yellow in Figure~\ref{subfig:verizon_combined}.
Table~\ref{tab: NCSH_CFCC} shows the number of census blocks where there is coverage according to the FCC but none according to Skyhook for each operator. Coverage claims in both tribal and non-tribal rural census blocks disagree the most. The number of such blocks are particularly high for Verizon (\rev{$17,994$}{$19,126$} overall) and T-Mobile (\rev{$17,389$}{$18,189$} overall).  There are two possible reasons for this disagreement: network operators lack \rev{}{adequate} infrastructure in rural areas, but tend to overestimate coverage while reporting it to FCC, or Skyhook is missing data points from rural census blocks where fewer people carry UEs. \rev{that detect the presence of LTE cells}{The latter case will lead to either some LTE cells not being detected or an inaccurate characterization of cell coverage due to fewer measurements.}

\begin{table}[]
\small
\begin{tabular}{|c|c|c|c|c|}
\hline
\textbf{Block type} & \textbf{Verizon} & \textbf{T-Mobile} & \textbf{AT\&T}  & \textbf{Sprint} \\ \hline \hline
% NT\_R & 528 (1\%)  & 2,635 (3\%) & 5,329 (6\%) & 20 (<1\%) \\ \hline
% NT\_U & 0 (0\%)    & 0 (0\%)    & 217 (1\%)  & 0 (0\%)  \\ \hline
% T\_R  & 2,655 (9\%) & 2,486 (8\%) & 2,548 (8\%) & 0 (0\%)  \\ \hline
% T\_U  & 0 (0\%)    & 0 (0\%)    & 2 (<1\%)    & 0 (0\%)  \\ \hline
\tiny{Non-Tribal Rural} & 528 (1\%)  & 2,575 (3\%) & 5,342 (6\%) & 19 (<1\%) \\ \hline
\tiny{Non-Tribal Urban} & 0 (0\%)    & 0 (0\%)    & 213 (1\%)  & 0 (0\%)  \\ \hline
\tiny{Tribal Rural}  & 2,655 (9\%) & 2,565 (8\%) & 2,166 (7\%) & 0 (0\%)  \\ \hline
\tiny{Tribal Urban}  & 0 (0\%)    & 0 (0\%)    & 4 (<1\%)    & 0 (0\%)  \\ \hline

\end{tabular}
\caption{\rev{}{Number of census blocks with LTE coverage according to the FCC, but only 3G coverage according to Skyhook. The numbers in parenthesis report the same data as a percentage of total census blocks of the corresponding type.}}
\label{tab: CSH_CELL_NCFCC}
%\vspace{-8mm}
\end{table}

To understand which of these potential reasons for disagreement is more likely, we check whether Skyhook shows 3G coverage for these census blocks (where the FCC reports LTE coverage but Skyhook does not). If Skyhook  reports 3G coverage in these blocks, this \rev{confirms}{suggests} that users \rev{}{may} have contributed to the Skyhook dataset in these census blocks, therefore LTE coverage would have been detected if it existed. \rev{}{Note that a more accurate way would have been to directly consider the location of end-user measurements connected using 3G technology and analyze whether they fall within LTE coverage areas in the FCC data. However, we did not have access to these end-user measurements due to Skyhook's privacy policy. Instead, we consider the 3G coverage maps as a reasonable approximation for our analysis and} generate a 3G coverage map at the census block level for these areas in the same manner as described previously for LTE. %The dataset and methodology used for generating the 3G coverage map is similar to the one used for generating the LTE coverage map. Using this \ignoreme{3G coverage} map, 
The number of  census blocks that show only 3G coverage according to Skyhook is presented in Table~\ref{tab: CSH_CELL_NCFCC}. We observe a significant number of census blocks where Skyhook detects 3G coverage, indicating \rev{that in these areas it is likely the FCC LTE coverage claims are overstated.}{that the FCC LTE coverage claims may be overstated in these areas.} The number of such blocks is greater for tribal rural areas (up to $9\%$), thus indicating a higher mismatch of the two datasets in tribal rural areas. 
\subsubsection{Active measurements compared to FCC and Skyhook coverage} \label{sec:ground_truth}
In this section, we compare our own active measurements with the coverage maps from the FCC and Skyhook described in Section~\ref{subsubsec:coverage_s_f}. We focus now on the geographic region around Santa Clara Pueblo, which lies north of Santa Fe (see Figure~\ref{fig: ue_drive}), a region with a mix of urban, rural, and tribal population blocks. 
% Santa Clara Pueblo should be capitalized, but the pueblos in the Pueblo area are not. 

%\vspace*{-0.05in}
\noindent\textbf{Methodology}: We use the \textit{Service State} readings collected in our measurements for this analysis (see Section~\ref{subsec:active_measurements}). %\textit{Service State} is a discrete variable indicating whether the phone is connected to a cell. 
We also collected information about the connected cell's technology (e.g. LTE) and the geolocation of the measurements. This information is used to infer whether LTE coverage exists at a location. We consider LTE to be available if the \textit{Service State} shows IN\_SERVICE to indicate an active connection, and if the associated cell is an LTE cell. We term this the \emph{active} LTE coverage. We then compare the FCC and Skyhook coverage with the active LTE coverage to see whether the datasets agree. Note that we use the coverage shapefiles for both Skyhook and the FCC in this comparison instead of the census block centroid approach in Section~\ref{subsubsec:coverage_s_f}. This allows us to compare coverage more precisely for a location, especially if a census block is only partially covered.

\begin{table}[t]
\small
\begin{subtable}{0.5\textwidth}
\centering
\begin{tabular}{|c|c|c|c|c|l|}
\hline
\multirow{2}{*}{\begin{tabular}[c]{@{}c@{}}\textbf{Active} \\ \end{tabular}} & \multirow{2}{*}{\begin{tabular}[c]{@{}c@{}}\textbf{Total}\\\end{tabular}} & \multicolumn{2}{c|}{\textbf{FCC}} & \multicolumn{2}{c|}{\textbf{Skyhook}} \\ \cline{3-6} 
                                                                         &                                                                               & \textbf{NC}         & \textbf{C}           & \textbf{NC}            & \textbf{C}            \\ \hline
                                                                         \hline
No Coverage (NC)                                                                       & 266                                                                           & 19\%     & 81\%      & 32\%        & 68\%       \\ \hline
Coverage (C)                                                                        & 1,440                                                                          & 0\%      & 100\%     & 5\%         & 95\%       \\ \hline
\end{tabular}
\caption{Verizon}
\label{tab:verizon_active}
\end{subtable}
%\vspace{-5pt}

\begin{subtable}{0.5\textwidth}
\centering
\begin{tabular}{|c|c|c|c|c|l|}
\hline
\multirow{2}{*}{\begin{tabular}[c]{@{}c@{}}\textbf{Active} \\ \end{tabular}} & \multirow{2}{*}{\begin{tabular}[c]{@{}c@{}}\textbf{Total} \\\end{tabular}} & \multicolumn{2}{c|}{\textbf{FCC}} & \multicolumn{2}{c|}{\textbf{Skyhook}} \\ \cline{3-6} 
                                                                         &                                                                                & \textbf{NC}         & \textbf{C}           & \textbf{NC}            & \textbf{C}            \\ \hline \hline
No Coverage (NC)                                                                       & 324                                                                            & 6\%      & 94\%      & 21\%        & 79\%       \\ \hline
Coverage (C)                                                                        & 1,361                                                                           & 0\%      & 100\%     & 5\%         & 95\%       \\ \hline
\end{tabular}
\caption{T-Mobile}
\label{tab:tmobile_active}
\end{subtable}
%\vspace{-5pt}

\begin{subtable}{0.5\textwidth}
\centering
\begin{tabular}{|c|c|c|c|c|l|}
\hline
\multirow{2}{*}{\begin{tabular}[c]{@{}c@{}}\textbf{Active}\\\end{tabular}} & \multirow{2}{*}{\begin{tabular}[c]{@{}c@{}}\textbf{Total}\\ \end{tabular}} & \multicolumn{2}{c|}{\textbf{FCC}} & \multicolumn{2}{c|}{\textbf{Skyhook}} \\ \cline{3-6} 
                                                                         &                                                                               & \textbf{NC}          & \textbf{C}          & \textbf{NC}            & \textbf{C}            \\ \hline \hline
No Coverage (NC)                                                                       & 568                                                                           & 25\%      & 75\%     & 53\%        & 48\%       \\ \hline
Coverage (C)                                                                        & 1,095                                                                          & 2\%       & 98\%     & 7\%         & 93\%       \\ \hline
\end{tabular}
\caption{AT\&T}
\label{tab:att_active}
\end{subtable}
%\vspace{-5pt}

\begin{subtable}{0.5\textwidth}
\centering
\begin{tabular}{|c|c|c|c|c|l|}
\hline
\multirow{2}{*}{\begin{tabular}[c]{@{}c@{}}\textbf{Active}\\ \end{tabular}} & \multirow{2}{*}{\begin{tabular}[c]{@{}c@{}}\textbf{Total} \\ \end{tabular}} & \multicolumn{2}{c|}{\textbf{FCC}} & \multicolumn{2}{c|}{\textbf{Skyhook}} \\ \cline{3-6} 
          &        & \textbf{NC} & \textbf{C} & \textbf{NC}   & \textbf{C}   \\ \hline \hline
No Coverage (NC)        & 231            & 96\%      & 4\%      & 99\%        & 2\%        \\ \hline
Coverage (C)         & 1,122            & 21\%      & 79\%     & 20\%        & 80\%       \\ \hline
\end{tabular}
\caption{Sprint}
\label{tab:sprint_active}
\end{subtable}

\caption{Confusion matrices comparing active measurement coverage with FCC and Skyhook. \textit{Total} denotes the number of active measurements in each category.}
\label{tab:active}
%\vspace{-10mm}
\end{table}

\begin{comment}
\begin{table}[]
\begin{tabular}{|c|c|c|c|c|c|c|c|}
\hline
\multirow{2}{*}{\textbf{Provider}} & \multirow{2}{*}{\textbf{Total}} & \multicolumn{2}{c|}{\textbf{Accuracy}} & \multicolumn{2}{c|}{\textbf{Precision}} & \multicolumn{2}{c|}{\textbf{Recall}} \\ \cline{3-8} 
                                   &                                 & \textbf{FCC}       & \textbf{SH}       & \textbf{FCC}        & \textbf{SH}       & \textbf{FCC}      & \textbf{SH}      \\ \hline
Verizon                            & 1706                            & 87\%               & 85\%              & 100\%               & 54\%              & 19\%              & 32\%             \\ \hline
T-Mobile                           & 1685                            & 82\%               & 81\%              & 100\%               & 50\%              & 6\%               & 21\%             \\ \hline
AT\&T                              & 1663                            & 73\%               & 79\%              & 87\%                & 80\%              & 25\%              & 52\%             \\ \hline
Sprint                             & 1353                            & 82\%               & 83\%              & 48\%                & 50\%              & 96\%              & 98\%             \\ \hline
\end{tabular}
\caption{Comparison of FCC and Skyhook (SH) coverage with our on-ground measurements.}
\label{tab:fcc_sh_gt}
\vspace{-3em}
\end{table}
\end{comment}

\noindent\textbf{Results}: Table~\ref{tab:active} shows the confusion matrices that compare active LTE coverage with reported coverage from the FCC and Skyhook maps. Both maps show coverage at locations where our measurements did not. In the case of Verizon, $81\%$ of the measurements with no coverage are from  locations reported as covered by the FCC. This over-reporting is lowest for Sprint and highest for T-Mobile. 

We also observe significant disagreement (up to $79\%$) between Skyhook coverage and our measurements. Two possibilities may cause this: i) paucity in Skyhook UE signal strength readings available for cell location and coverage radius estimation, or ii) error in the cell propagation model itself possibly due to variations in the environment conditions such as the terrain. In either case, Skyhook agrees better with our  measurements than the FCC in reporting areas with no LTE coverage. E.g., in the case of AT\&T, $75\%$ of our measurements with no LTE coverage belong to areas reported as covered by the FCC as compared to $48\%$ by Skyhook. 

\section{Recommendations}% and Future directions} 
\label{sec:discussion}
In this section, we discuss some of the implications of our experience collecting and analyzing coverage data, recommendations based on our findings, and directions for future work.

\textbf{Recommendations for the FCC}: Our findings make a case for including mechanisms that validate ISP-reported coverage data, especially in rural and tribal regions. Given the scale of cellular networks, crowdsourcing coverage measurements is a viable approach to validate access as opposed to controlled measurements. Within crowdsourcing, we suggest leveraging \textit{incidental} rather than \textit{voluntary} approaches, possibly working with third-party services that collect network measurements as part of their service process (as in  the case of Skyhook). 

In addition, crowdsourcing alone may not be sufficient for determining coverage in some cases. Even with the more complete datasets provided through incidental crowdsourcing, rural areas tended to receive significantly fewer measurements per tower. In such cases, mechanisms need to be developed to precisely determine areas of greatest disagreement using sparse crowdsourced datasets. Resources can then be focused to target data collection in these areas instead of a blanket approach measuring coverage everywhere.

\textbf{Recommendations for crowdsourced data collection}: We find some shortcomings in the existing crowdsourced datasets. First, existing datasets only report areas with positive coverage, i.e., areas where coverage is observed. This makes it difficult to distinguish areas that lack coverage from areas for which no measurements were gathered. Recording areas that lack a usable signal can enable more stronger conclusions from crowdsourced data. 

Second, we note that even crowdsourced datasets are prone to overestimation of coverage potentially due to errors in cell location and coverage estimation. Research efforts that effectively utilize the knowledge of cellular network design are needed for an accurate characterization of coverage from crowdsourced measurements. For instance, \rev{a single physical tower in an LTE network hosts multiple cells each serving different sectors operating under different frequencies. One can use this fact to design algorithms that consider combined localization of multiple proximate cells. This is in contrast to existing techniques that localize cells independently that can lead to higher errors with fewer end-user measurement.}{existing cell location estimation techniques localize cells independently (see Section~\ref{sec:skyhook_data}) and are prone to errors when there are few end-user measurements~\cite{li2017identifying}. Instead, one can utilize the fact that a single physical tower in an LTE network hosts multiple cells. Thus, algorithms that jointly localize cells for whom the end-user measurements are in physical proximity may provide higher accuracy even with fewer end-user measurements.} Similarly, alternate data sources can also be considered for localizing cell infrastructure such as using geo-imagery data to identify physical towers or directly obtaining infrastructure data from entities that build and manage physical cell towers (usually different from cellular ISPs).

\textbf{Measuring access beyond binary coverage}: While the focus of this work is on understanding coverage, we recognize that a binary notion of coverage alone does not necessarily indicate the existence of usable LTE connectivity. Various other factors can impact end-user experience in a ``covered" area such as low signal strength or \textit{poor} middle-mile connectivity. Thus, future coverage measurement efforts need to augment coverage reports with measurements of performance to provide models that are more aligned with user experiences. Measuring such performance metrics poses a greater challenge because end-user experience depends on a myriad of factors beyond just last-mile link quality. We believe that efforts that lead to increased community awareness (e.g., workshops in public libraries, community meetings) on the importance of measuring mobile coverage is the way to tackle this problem.   

Finally, we also note that access and adoption are different and there are issues beyond access that might also warrant measurement and consideration as accountability measures for operators. Our collection of ground truth data sets involved five days driving through Rio Arriba County in northern New Mexico. In preparation for the trip, we worked to obtain SIM cards that would enable us to access the networks of the four major U.S. LTE operators. This was surprisingly difficult; over the course of a month leading up to the measurement campaign, we spent a collective 24 hours in various operator kiosks and stores in three states in order to obtain four SIM cards (one for each major operator). At one of the stores in Santa Fe, we encountered a woman who had to drive an hour from Las Vegas, NM to address some of the issues she was having with her mobile service operator that were preventing her from using her data plan. While these anecdotal experiences mirror the qualitative claims of coverage overestimation, they do introduce a new set of issues that need to be taken into account to effectively reduce the barriers of Internet access for rural communities.

\section{Conclusion}
\label{sec:conclusion}
% Our work draws the attention of the computing community towards the limitations of the existing coverage datasets and the broader challenges in understanding mobile broadband access 
In this paper, we quantitatively examine the LTE coverage disagreement among existing datasets collected using different methodologies. We find that existing datasets display the most divergence when compared with each other in rural and tribal areas. We discuss our findings with respect to their implications for telecommunications policy. % Critically, we recommend that while incidental crowdsourced datasets can be used to validate coverage reports generated by providers, in rural and tribal areas, disagreement between datasets can be leveraged to identify spaces where more concentrated validation efforts are needed. 
We also identify several future research directions for the computing community, including: mechanisms to augment existing datasets to precisely determine areas where more concerted measurement efforts are needed,  improved coverage estimation models especially for areas with a lower density of crowdsourced measurements, and accurate and scalable measurement of access beyond a binary notion of coverage. 

\section*{Acknowledgements}
This work is funded in part by National Science Foundation Smart and Connected Communities grant NSF-1831698.

\bibliographystyle{ACM-Reference-Format}
\bibliography{ref.bib}

\end{document}